\documentclass[aps,prb,twocolumn,amsmath,amssymb,showpacs,superscriptaddress]{revtex4-1}

\usepackage{graphicx}
\usepackage{dcolumn}
\usepackage{bm}

\def\tio{TiO$_2$}
\def\ra{\rightarrow}
\def\phm#1{\phantom{#1}}
\def\beq{\begin{equation}}
\def\eeq{\end{equation}}
\def\R{\mathbf{R}}

\def\rb{\mathbf{r}}
\def\dr{\mathrm{d}}
\def\lr{\mathrm{lr}}
\def\sr{\mathrm{sr}}
\def\wt#1{\widetilde{#1}}

\begin{document}

\title{First-principles DFT+GW study of oxygen vacancies in rutile \tio}

\author{Andrei Malashevich}
\affiliation{
Department of Physics, University of California,
Berkeley, California 94720, USA}
\affiliation{
Materials Sciences Division, Lawrence Berkeley National Laboratory,
Berkeley, California 94720, USA
}
\affiliation{
Department of Applied Physics, Yale University,
New Haven, Connecticut 06511, USA
}
\author{Manish Jain}
\affiliation{
Department of Physics, University of California,
Berkeley, California 94720, USA}
\affiliation{
Materials Sciences Division, Lawrence Berkeley National Laboratory,
Berkeley, California 94720, USA
}
\affiliation{
Department of Physics, Indian Institute of Science, 
Bangalore 560012, India
}
\author{Steven G. Louie}
\email{sglouie@berkeley.edu}
\affiliation{
Department of Physics, University of California,
Berkeley, California 94720, USA}
\affiliation{
Materials Sciences Division, Lawrence Berkeley National Laboratory,
Berkeley, California 94720, USA
}

\date{\today}

\begin{abstract}
We perform first-principles calculations of the quasiparticle defect states,
charge transition levels, and formation energies of oxygen vacancies in 
rutile titanium dioxide. The calculations are done within the recently developed
combined DFT+GW formalism, including the necessary electrostatic corrections
for the supercells with charged defects.
We find the oxygen vacancy to be a negative $U$ defect, where $U$ is the defect
electron addition energy. For the values of Fermi level below $\sim2.8$~eV
(relative to the valence band maximum) we find the $+2$ charge state 
of the vacancy to be the most stable,
while above $2.8$~eV we find that the neutral charge state is the most stable.
\end{abstract}

%
%

\pacs{61.72.jd,61.72.Bb,71.20.-b,71.18.+y}
\maketitle

\section{Introduction}
\label{sec:intro}

Titanium dioxide (\tio) attracts a lot of attention of researchers as
a versatile functional material used in numerous technological applications
including photocatalysis, hydrolysis, solar cells, high-$k$ dielectrics,
optoelectronic devices, sensors, etc.
\cite{grant_59,diebold_03,augustynski_93,linsebigler_95,oregan_91,yang_08,kim_08,
wilk_01,fujishima_72} Lattice defects, such as vacancies, substitution impurities,
and interstitial impurities, inevitably occur in materials regardless of whether they
are synthesized or created naturally. These defects can greatly influence
the mechanical, electrical, thermal, and optical properties of solids.

\begin{figure}
\centering\includegraphics[height=4cm]{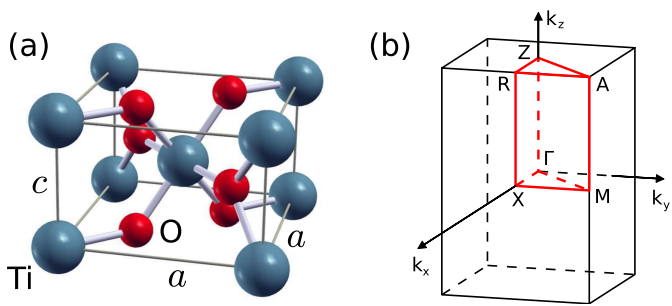}
\caption{(Color online.) (a) Tetragonal primitive cell of \tio\
  in the rutile crystal structure. (b) The corresponding Brillouin zone.
}
\label{fig:tio2}
\end{figure}

Among the major three crystal polymorphs of \tio, rutile is the most common one,
the other two being anatase and brookite. Rutile \tio\ has a tetragonal
primitive cell with two formula units (see Fig.~\ref{fig:tio2})
and its symmetry is described by the space group $P4_2/mnm$. 
The lattice parameters are $a=4.594$~\AA\ and $c=2.959$~\AA\ 
at room temperature.
The Ti and O atoms reside at the $2a$ and $4f$ Wyckoff positions, the latter
characterized by the single internal parameter $u=0.305$.\cite{abrahams_71}

Rutile \tio\ in its stoichiometric form is an insulator with an optical band gap
of $3.0$~eV.\cite{cronemeyer_59,amtout_95} The optical gap, however,
is smaller than the electronic band gap due to electron-hole interactions.
The latter band gap is connected to a single-particle (or quasiparticle)
description and can be measured in photoemission experiments. The values for
the electronic gap in the literature vary in the range of $3.3-4.0$~eV. 
\cite{tezuka_94,hardman_94,rangan_10} 
For more discussion on the relation between the electronic and
optical gaps of \tio, as well as comparison of experimental and theoretical values,
see, e.g., Ref.~\onlinecite{chiodo_10}.

Heating rutile crystals in reducing atmosphere
results in the increase of the ($n$-type) electrical conductivity and
rutile composition changes to non-stoichiometric TiO$_{2-x}$.
This change is attributed to various types of defects such as oxygen vacancies,
Ti$^{3+}$ and Ti$^{4+}$ interstitials, and planar defects.\cite{diebold_03}

In this work we perform calculations of the charge transition levels
and defect formation energies of oxygen vacancies in three charge states
following a recently developed DFT+GW approach.\cite{hedstrom_06,*rinke_09,jain_11}
  There are several advantages of this approach over the traditional DFT-only
  approaches.\cite{iddir_07} In this method, the GW correction of DFT eigenvalues takes care
  of the self-energy and self-interaction errors and resolves the problem
  of band-gap underestimation. The latter problem is often responsible 
  for incorrect DFT prediction of defect level position outside the bulk band gap.
  In addition, within the DFT+GW approach the formation energies
  can be calculated without computing differences in total energies 
  of systems with {\it different} number of electrons.
We note that the need to go beyond the standard DFT approaches for calculations
of defect formation energies and charge transition levels
is now well recognized, as more studies of the electronic structure
of oxides based on hybrid functionals and GW perturbation methods
appear in the literature. Recently, Peng and collaborators\cite{peng_13}
proposed an alternative scheme to calculate defect formation energies
by using GW to correct band edge energies.

The rest of the paper is organized as follows. Section \ref{sec:methods}
describes in detail the DFT+GW formalism used in our calculations.
Computational details are given in section \ref{sec:computational}.
Then the main results are presented in section \ref{sec:results}, followed
by a summary in section \ref{sec:summary}.

\section{Methods}
\label{sec:methods}

\subsection{DFT+GW formalism}
\label{subsec:formalism}

The DFT+GW approach employed in this work is developed
in Refs.~\onlinecite{hedstrom_06,*rinke_09} and
\onlinecite{jain_11}.
Here, we will introduce the notations used in the subsequent sections.

We describe the atomic state of a system with defect in a charge state $q$ 
(oxygen vacancy $V_{\mathrm{O}}^q$ in our case)
by a generalized coordinate $\R$. In general, $\R$ corresponds to an arbitrary
configuration, not necessarily equilibrium configuration. The equilibrium
configuration of the defect in a charge state $q$ we will denote as $\R_q$.
One can define\cite{jain_11}
the defect formation energy $E^{\mathrm{f}}_q(\R,\mu_{\mathrm{O}},E_{\mathrm{F}})$,
which depends on the chemical potential of oxygen  $\mu_{\mathrm{O}}$
(determined by the experimental preparation conditions)
and the Fermi level $E_{\mathrm{F}}$. We reference $E_{\mathrm{F}}$
to the valence band maximum (VBM), so it can take values between
zero and the bulk band gap depending on the specific sample.

Charge transition level $\epsilon^{q/q-1}$ is defined as the Fermi level 
at which the charge state of the defect changes from $q$ to $q-1$
or, in other words, at which the formation energies of the defect
in charge states $q$ and $q-1$ are equal. One can show that
the value of the charge transition level can be separated into two
contributions as $\epsilon^{q/q-1}=E_{\mathrm{relax}} + E_{\mathrm{QP}}$,
where $E_{\mathrm{QP}}$ is a quasiparticle excitation energy (addition
or removal of a single electron) and $E_{\mathrm{relax}}$ is 
the (atomic) relaxation energy of the defect in the new charge state.
Since $E_{\mathrm{relax}}$ is given
by the difference in the total energies of the system whose 
total number of electrons remains unchanged, it can be calculated accurately
using standard DFT methods, while $E_{\mathrm{QP}}$ may be evaluated
using the {\it ab initio} GW method.\cite{hybertsen_86}

The combined DFT+GW approach avoids the typical problems
one encounters when using DFT for all terms,
such as the underestimation of the band gap and self-interaction errors.

\subsection{Electrostatic corrections}

Ideally, when studying defects, one would like to consider a single defect
in an infinite bulk material. In practice, however, one often uses a supercell 
approach,\cite{cohen_75}
in which a finite supercell with defect is constructed and periodic
boundary conditions are applied. If the supercell is not large enough
the spurious interactions between the defect and its own images should be taken
into account. For charged defects, in particular, the spurious long-range
Coulomb potential from defect images results in a shift of the defect state
in the bulk band gap.
This effect has been shown to be quite significant
for oxygen vacancies in hafnia.\cite{jain_11}

There are several ways to calculate the electrostatic corrections,
to be denoted as
$\Delta{E}_{\mathrm{QP}}^{\mathrm{e.s.}}$. All of them can be done
withing the DFT-only formalism since the spurious potential is electrostatic 
and affects only the Hartree potential in the DFT calculation. Further, Hartree
potential is not affected by the self-energy operator within our GW approach. 
The straightforward approach would be to increase the size of the 
supercell with defect and keep track of the shift in the Kohn-Sham eigenvalue
corresponding to the defect state. Taking into account the fact that the 
strength of the Coulomb interaction is inversely proportional to the distance,
one can extrapolate the change in the Kohn-Sham eigenvalue to infinite
supercell size.\cite{jain_11} This approach, however, requires construction of
supercells with very large number of atoms
(often thousands of atoms are required).

In this work, we opted for a different approach proposed by
Freysoldt and collaborators,\cite{freysoldt_09}
which does not require a construction of extremely large supercells. 
The only requirement on the supercell size is that the charge density
associated with the defect state is well localized in a small volume
inside the supercell.
In the following, we describe the main changes to this method 
adapting it to DFT+GW framework. We shall keep the original
notations and definitions.

If a neutral defect state can be described by a local wavefunction $\psi_{\dr}$ 
then one can calculate the {\it unscreened} charge density $q_{\dr}(\rb)$ associated 
with the {\it charged} defect (assuming the charge $q$ goes entirely 
to the local defect state).
The charge $q$ then becomes screened by the surrounding electrons.
The corresponding change in the electrostatic potential relative
to the neutral defect is denoted by $V_{q/0}$. Note that
in this discussion, as in the original formulation,\cite{freysoldt_09} we do
not consider effects of lattice relaxations due to the change of the charge
state of the defect.

Now we consider a periodic system corresponding to an array of charged
defects and add a compensating homogeneous background charge with density
$n=-q/\Omega$, where $\Omega$ is the supercell volume.
Assuming a linear-response behavior, the change in the electrostatic
potential for this system $\wt{V}_{q/0}(\rb)$ is given by a superposition
of the potentials $V_{q/0}(\rb+\R)$ up to a constant, where $\R$ 
denotes lattice vectors.
Thus, knowing $V_{q/0}(\rb)$ of an infinite system one can reproduce
the potential $\wt{V}_{q/0}(\rb)$ of a periodic system (up to a constant).
The spurious electrostatic potential induced by the images of the defect
in the home supercell is, thus, given by $[\wt{V}_{q/0}(\rb)-V_{q/0}(\rb)]$.
Within DFT, this corresponds to the undesired shift of the Kohn-Sham defect state
\beq
\label{eq:shift}
\Delta\epsilon^{\mathrm{KS}}_{\dr}=
-\int_{\Omega}d^3r\,\vert\psi_{\dr}(\rb)\vert^2[\wt{V}_{q/0}(\rb)-V_{q/0}(\rb)].
\eeq

In practice, we can compute the periodic potential $\wt{V}_{q/0}(\rb)$
but we do not know the original potential $V_{q/0}(\rb)$ of the infinite system.
At large distances this potential may be well approximated by the long-range
screened Coulomb potential\cite{freysoldt_09} $V_{q/0}^{\lr}(\rb)$,
which requires knowledge
of the dielectric constant $\varepsilon$ (which, in turn,
can be found, e.g., from density-functional perturbation theory) for its evaluation.
Thus, the idea is to separate the potential $V_{q/0}(\rb)$ into long-range
and short-range parts as $V_{q/0}(\rb)=V_{q/0}^{\lr}(\rb)+V_{q/0}^{\sr}(\rb)$.
Assuming that the short-range potential decays rapidly with distance
and is essentially zero at the border of the supercell (with defect
placed in the center of the supercell),
we can write for $\rb\in\Omega$
\beq
\label{eq:V_sr}
\wt{V}_{q/0}^{\sr}(\rb)=V_{q/0}^{\sr}(\rb)+C,
\eeq
where the constant $C$ absorbs the ambiguity in the absolute position of
$\wt{V}_{q/0}$. This constant may be found by requiring that 
$\wt{V}_{q/0}$ and $\wt{V}_{q/0}^{\lr}$ align far from the defect.

Hence, the shift of the defect state due to the spurious electrostatic
potential, Eq.~(\ref{eq:shift}), can be calculated from two parts,
each coming from the long-range and short-range contributions to the potential.
The first part is given by
\beq
\label{eq:shift_lr}
\Delta\epsilon^{\mathrm{KS}}_{\dr,\lr}=
-\int_{\Omega}d^3r\,\vert\psi_{\dr}(\rb)\vert^2[\wt{V}_q^{\lr}(\rb)-V_q^{\lr}(\rb)]
\eeq
and the second part is given by
\beq
\label{eq:shift_sr}
\Delta\epsilon^{\mathrm{KS}}_{\dr,\sr}=
-\int_{\Omega}d^3r\,\vert\psi_{\dr}(\rb)\vert^2
[\wt{V}_{q/0}^{\sr}(\rb)-V_{q/0}^{\sr}(\rb)]
=-C.
\eeq
Equations (\ref{eq:shift_lr}) and (\ref{eq:shift_sr}) give the spurious {\em shift}
of the Kohn-Sham level, while the electrostatic correction
$\Delta{E}^{\mathrm{e.s.}}_{\mathrm{QP}}$ that needs to be applied is 
\beq
\label{eq:es_cor}
\Delta{E}^{\mathrm{e.s.}}_{\mathrm{QP}}=-\Delta\epsilon^{\mathrm{KS}}_{\dr,\lr}-
\Delta\epsilon^{\mathrm{KS}}_{\dr,\sr}.
\eeq

  To see how the described above method works, we performed a calculation
  of the oxygen vacancy in rock-salt MgO in its $+1$ charge state.
  For simplicity, we performed a spin unpolarized calculation using $2\times2\times2$
  cubic supercelli (63 atoms). We found a Kohn-Sham eigenvalue in the bulk band gap 
  corresponding to a defect state located $1.07$~eV above the VBM.
  The electrostatic correction calculated with the above method 
  resulted in a shift of defect eigenvalue of $-0.65$~eV, where $-0.45$~eV
  comes from the first term in Eq.~(\ref{eq:es_cor}) and $-0.20$~eV
  comes from the second term.
  Then, we performed calculations using $3\times3\times3$ (215 atoms)
  and $4\times4\times4$ (511 atoms) supercells. We found the defect eigenvalue
  to be $0.83$~eV and $0.64$~eV above VBM in 215-atom and 511-atom supercells,
  respectively. We fit the defect eigenvalue to $\epsilon_d=\epsilon_d^0+A/(L)$,
  where $L$ is the size of the supercell in arbitrary units (e.g., $L=2,3,4$ in our case),
  $\epsilon_d^0$, and $A$ are fitting parameters. This way, we found, in the limit of
  infinite supercell, the electrostatic correction to be $-0.84$~eV in a reasonable agreement
  with the previous result.

  Recently, a similar procedure for correcting the Kohn-Sham eigenvalues
  due to electrostatic spurious potential was suggested by Chen and Pasquarello.
  \cite{chen_13}

\section{Computational details}
\label{sec:computational}
In this work all mean field calculations were done within the 
density functional theory (DFT) framework.
It has been shown recently that structural relaxation in the case of rutile
\tio\ depends strongly on the choice of
exchange-correlation potential.\cite{janotti_10}
Adequate description of the crystal structure can be 
obtained using hybrid functionals, such as that of Heyd, Scuseria,
and Ernzerhof (HSE).\cite{heyd_03,*heyd_06}
If the crystal structure of rutile \tio\ with oxygen vacancy
is relaxed, e.g., using the Perdew, Burke, and Ernzerhof (PBE)
exchange-correlation potential,\cite{perdew_96}
the defect level moves into conduction band regardless
of its charge state.\cite{janotti_10}
For this reason, in our work, all structural relaxations
(both for bulk \tio\ and supercells with defects) were performed using HSE06
hybrid functional,\cite{heyd_03,*heyd_06} in which 25\% of the (short-range)
Hartree-Fock (HF) exchange is mixed with 75\% PBE exchange.
We used projector augmented-wave (PAW) method
\cite{blochl_94,kresse_99}
as implemented in the VASP code package.\cite{VASP1,*VASP2}
The standard PBE pseudopotentials for both Ti and O supplied with the VASP
package were employed. For Ti, the $3s$, $3p$, $3d$, and $4s$ states were 
treated as valence orbitals. We used a plane-wave basis set with an energy cut-off
of 450~eV.

For bulk \tio, the Brillouin zone was sampled by a uniform $4\times4\times6$
$k$-point mesh. Oxygen vacancies were simulated by constructing a 
$2\times2\times3$ supercell of 72 atoms and removing one O atom.
Brillouin zone integrations for the supercells were performed using
an equivalent $2\times2\times2$ mesh of $k$ points.

Once the structural parameters for a system of interest were determined, 
we performed a separate
self-consistent field (SCF) calculation using PBE exchange-correlation
potential
in order to obtain a mean-field starting point for our GW calculations.
For this purpose we used {\sc Quantum ESPRESSO} code package.\cite{QE_09} 
Troullier-Martins norm-conserving pseudopotentials\cite{troullier-prb91}
were generated for Ti and O.
For Ti, the $3s$ and $3p$ semi-core states were treated as valence and
the pseudopotential was generated in the Ti$^{4+}$ configuration.
The cut-off radii for the $3s$, $3p$, and $3d$ states were chosen
to be $0.9$, $0.9$, and $1.0$ a.u., respectively. The energy cut-off
for the plane-wave basis of 200~Ry was used in this case.

The GW calculations were performed using the BerkeleyGW code
package.\cite{hybertsen_86,deslippe_12}
We used a G$_0$W$_0$ approach within the complex generalized
plasmon-pole (GPP) model.\cite{zhang_89}
For the dielectric matrix calculation, the frequency cut-off
was chosen to be 40~Ry and the number of valence and conduction bands 
was chosen to be 2\,000 for bulk rutile \tio\ and 4\,000 for the supercell
calculations. In case of supercells, the convergence with respect
to empty states is not guaranteed despite the large number of states
used in our calculations. For this reason, the extrapolation to infinite
number of states is required. We used the static-remainder method
for this purpose.\cite{deslippe_13}

\section{Results}
\label{sec:results}
\subsection{Bulk rutile \tio}

Structural properties of bulk rutile \tio\ were calculated using both PBE 
and HSE06 exchange-correlation potentials. The results of these calculations
are in a very good agreement with each other and experiment as can be seen 
from Table~\ref{tab:struct}.

\begin{table}
\caption{\label{tab:struct}
Calculated and experimental structural parameters of rutile TiO$_2$.}
\begin{ruledtabular}
\begin{tabular}{lccc}
& $a$   & $c/a$ & $u$ \\
& (\AA) &       &     \\
\hline
PBE      & 4.64 & 0.639 & 0.305 \\
HSE06    & 4.58 & 0.646 & 0.305 \\
Expt.\footnotemark[1] & 4.59 & 0.644  & 0.305 \\ 
\end{tabular}
\footnotetext[1]{Ref.~\onlinecite{abrahams_71}.}
\end{ruledtabular}
\end{table}

\begin{figure}
  \centering\includegraphics[width=246pt]{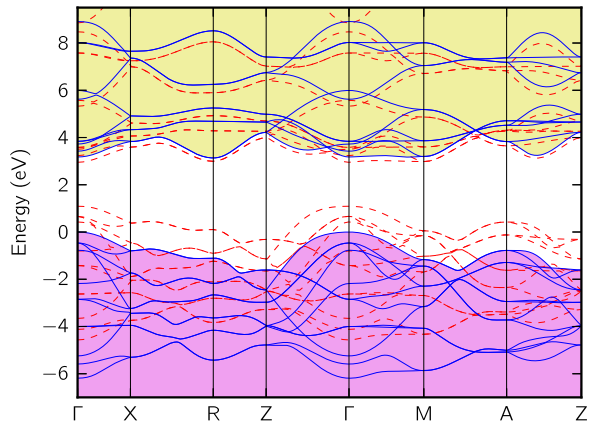}
  \caption{(Color online.)
      Theoretical band structure of rutile \tio\ calculated
        within DFT using the PBE exchange-correlation potential
          (red dotted lines) and using the GW method (blue solid lines).
        }
  \label{fig:bands}
\end{figure}

The electronic band structure was computed along high symmetry lines
[the labels for the high symmetry points in the Brillouin zone
are shown in Fig.~\ref{fig:tio2}~(b)].
The band structure plots before
and after the self-energy correction are shown in Fig.~\ref{fig:bands}.
As one can see from the figure, the effect of the G$_0$W$_0$ correction
to a first approximation can be considered as a scissor-shift operation,
although the corrections to some bands are larger than to the others.

Within PBE, the calculated band gap is a direct gap of only $1.86$~eV
at the $\Gamma$ point. 
After applying the GW correction, we found the fundamental
gap to be the indirect $\Gamma-\mathrm{R}$ gap of $3.13$~eV although the 
direct gap at the $\Gamma$ point of $3.18$~eV is very close
to the $\Gamma-\mathrm{R}$ gap.
A more detailed analysis of the band structure of bulk rutile \tio,
including the calculation of quasiparticle effective masses,
is given in the Supplemental Material.\cite{supplemental}

\subsection{Oxygen vacancy}

\begin{figure}
\centering\includegraphics[]{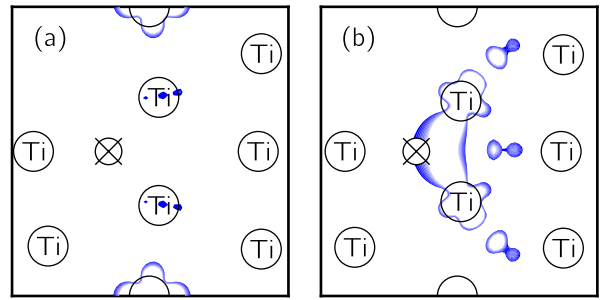}
\caption{ (Color online.) Charge densities of the localized states 
in the gap found in the HSE calculations with $+1$ charged supercells. Panel (a)
shows the ground state, corresponding to a $+2$ charged oxygen vacancy 
and a polaron. Panel (b) shows the $+1$ charged oxygen vacancy.
The isosurfaces show the $10\%$ of the charge densities
of the localized states.
}
\label{fig:hse-density}
\end{figure}

\begin{figure}
\centering\includegraphics[]{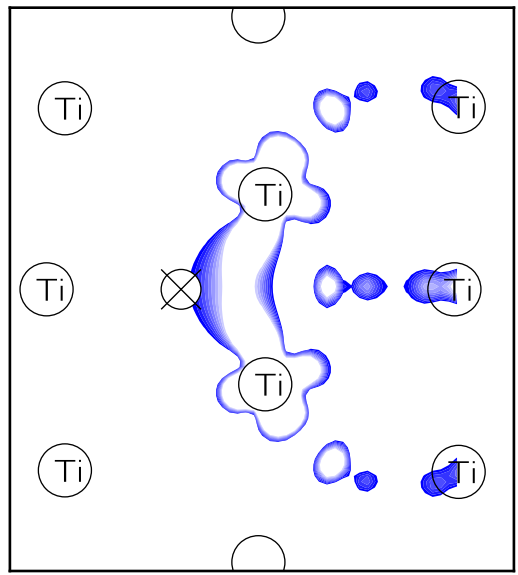}
\caption{ (Color online.) Charge density of the localized defect
  state of the $+1$ charged oxygen vacancy found in PBE calculation.
The isosurface shows the $10\%$ of the charge density.
The atomic positions are the same as in Fig.~\ref{fig:hse-density}~(b).
}
\label{fig:pbe-density}
\end{figure}

Qualitatively the important defect state in the gap
associated with the oxygen vacancy
in rutile \tio\ can be understood as follows. In bulk rutile,
each O atom is surrounded by three neighboring Ti atoms.
When one O atom is removed, the three Ti dangling bonds (mostly
having $d$ character) form a low-energy state of $a_1$ 
symmetry.\cite{janotti_10} In the neutral charge state of
the oxygen vacancy ($V_{\mathrm{O}}^0$), the defect state $a_1$
is doubly occupied.
In the $+1$ charge state ($V_{\mathrm{O}}^{+}$), 
the $a_1$ is singly occupied.
In both cases the occupied $a_1$ state bonds the
neighboring Ti atoms and keeps them from moving away from the vacancy.
In the $+2$ charge state ($V_{\mathrm{O}}^{2+}$), the $a_1$ state
is unoccupied, which results in a much larger displacements of the Ti atoms
outward from the vacancy site.

It is worth noting that if one simply relaxes the $0$ or $+1$ charged systems
within PBE, one may find a ground state which does not necessarily
correspond to the electrons bound to the vacancy site.
Recently, first-principles calculations\cite{deak_11,deak_12,janotti_13}
have shown that polarons may form in \tio.
In addition, experimental evidence of intrinsic polarons in rutile
has been seen in electron paramagnetic resonance measurements.\cite{yang_13}
Indeed, in our calculations we find that a na\"ive relaxation
of the $+1$ charged system leads to a ground state with
an electron away from the vacancy site.
We find a localized state with its eigenvalue in the gap,
but the charge density corresponding to this state is not localized
at the vacancy site but is localized at the next-nearest neighbor Ti atom.
While in principle a polaron can be formed anywhere in the supercell,
its localization on the next-nearest Ti atom can be attributed to
the finite size of the supercell used in our calculations.
Figure~\ref{fig:hse-density}~(a) shows the calculated charge density
of the state in the gap for such a polaron ground state.
However, for the purpose of calculation of charge transition levels,
this particular state is not appropriate. In order to stabilize the $+1$
vacancy state of interest (i.e., the electron bound to the vacancy site),
we used the following procedure. First, we performed a spin-unpolarized
relaxation of the neutral vacancy. This resulted in a state with two electrons
bound to the vacancy site. Second, we relaxed the $+1$ charged system starting
from the atomic configuration found in the first step.
This procedure ensured that the defect state remained bound to the vacancy site.
Figure~\ref{fig:hse-density}~(b) shows the charge density of the obtained
$+1$ vacancy defect state. We emphasize again that the state thus found is not a ground state
(i.e., lowest total energy) in our calculations but rather a local minimum.
We found that it is above the ground state (we call it a polaron ground state)
by $1.2$~eV.

For the purpose of doing the GW calculation, we used a PBE mean field solution
from a structure determined with HSE. This was done because GW calculation
requires a large number of empty bands and the computational cost of using
HSE as the mean field becomes prohibitive. This is a reasonable procedure
because GW is a perturbative correction and does not depend sensitively
on the starting mean field. Because our GW calculation is a 
G$_0$W$_0$ calculation, we ensured that the resulting PBE defect wavefunction
is similar to the one obtained from HSE. Figure \ref{fig:pbe-density}
shows the charge density from the defect wavefunction obtained within PBE.
Comparing this figure to the Fig.~\ref{fig:hse-density}~(b), we can see that
the defect state charge densities obtained using PBE and HSE for the same structure
are similar.

\begin{figure}
\centering\includegraphics[]{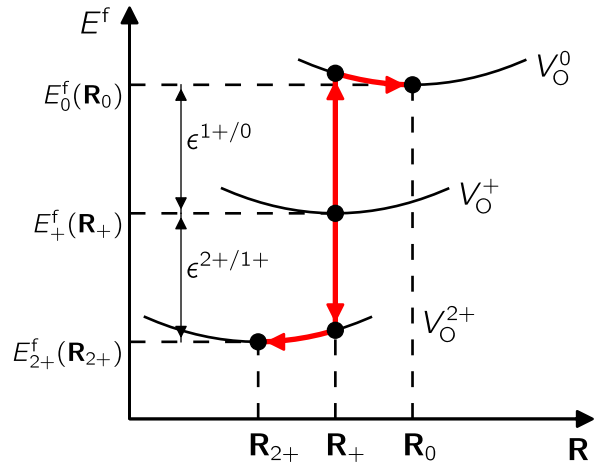}
\caption{Schematic illustration of the
  calculation of the charge transition levels $\epsilon^{2+/1+}$
  and $\epsilon^{1+/0}$ within the DFT+GW formalism.
  Arrows indicate the actual paths used in our calculations.
}
\label{fig:ctl_actual}
\end{figure}

In order to calculate the charge transition levels, we started from 
$+1$ charged oxygen vacancy. Figure~\ref{fig:ctl_actual} schematically
illustrates the paths in formation energy vs generalized coordinate space
that we took. It has been shown that all paths in this space
give the same value of charge transition
levels to within $\pm0.1$~eV provided that electrostatic corrections are taken
into account.\cite{jain_11} To reduce the computational cost,
we performed GW calculation on the $+1$ charged oxygen vacancy.
This allows us to calculate both $\epsilon^{1+/0}$ and $\epsilon^{2+/1+}$
as can be seen from Fig.~\ref{fig:ctl_actual}. For $\epsilon^{1+/0}$
we computed the quasiparticle (quasielectron) energy of the lowest unoccupied
localized state (which in our case turned out to be slightly above the CBM).
For $\epsilon^{2+/1+}$ we computed the quasiparticle (quasihole) energy
of the $+1$ defect state.
Both quasiparticle energies were evaluated relative to the
valence band maximum $E_\mathrm{v}$,
since we defined $\epsilon^{q/q-1}$ relative to
$E_\mathrm{v}$ in Sec.~\ref{subsec:formalism}.

Table \ref{tab:ctl} shows the results of our computed quasiparticle and relaxation
energies as well as the corresponding charge transition levels.
From the Table it is clear that the oxygen vacancies are negative $U$ defects,
where $U$ is the defect charging energy.
Also from the Table, one can see that electrostatic corrections are not negligible
and have to be included into the calculation.

Further, one can calculate the absolute formation energies
as a function of Fermi energy. For a given chemical potential of oxygen,
one needs to know the formation energy of the neutral vacancy,
which can be calculated within DFT, since for $q=0$ the absolute
values of Kohn-Sham levels do not enter in the definition of formation
energy.\cite{jain_11}
Note also that formation energy
of the neutral vacancy does not depend on the value of Fermi level 
$E_{\mathrm{F}}$.
Then using the definition of charge transition levels, 
one can obtain
the formation energy for all the charge states for a given chemical
potential of oxygen. Namely, for a $+1$ oxygen vacancy $V_{\mathrm{O}}^{+}$
one can write
\beq
\label{eq:form_en_1}
E^{\mathrm{f}}_{1+}(E_{\mathrm{F}})=
E^{\mathrm{f}}_0-\epsilon^{1+/0}+E_{\mathrm{F}},
\eeq
while a corresponding relation for $V_{\mathrm{O}}^{2+}$ is
\beq
\label{eq:form_en_2}
E^{\mathrm{f}}_{2+}(E_{\mathrm{F}})=
E^{\mathrm{f}}_0-\epsilon^{1+/0}-\epsilon^{2+/1+}+2E_{\mathrm{F}}.
\eeq
It is worth noting that calculating formation
energies of the charged defects in this manner does not involve
the value of the valence band maximum within mean field.
This ensures that the energy scale for the electrons is set only
by the GW calculation and not by DFT calculations. 

\begin{figure}
\centering\includegraphics[]{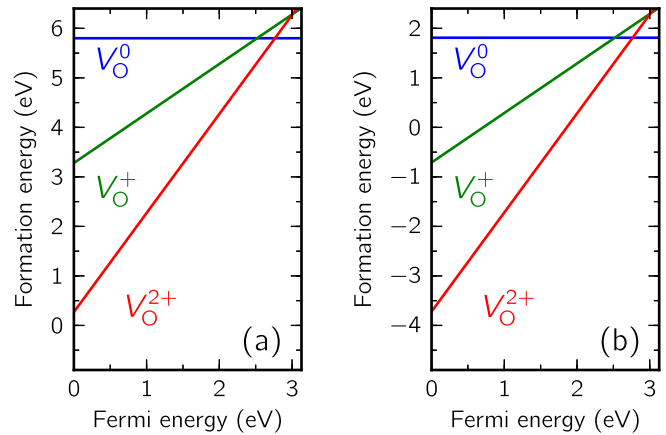}
\caption{ Calculated formation energies of oxygen vacancies in rutile
  \tio\ plotted as functions of Fermi level $E_{\mathrm{F}}$ in the
  (a) oxygen-rich and (b) titanium-rich growth conditions.
}
\label{fig:form_en}
\end{figure}

Figure \ref{fig:form_en} shows our results for formation energy of various
charges states of the oxygen vacancy plotted as a function of Fermi energy
$E_{\mathrm{F}}$ in the oxygen-rich
[Fig.~\ref{fig:form_en}~(a)] and oxygen-poor [Fig.~\ref{fig:form_en}~(b)]
growth conditions. For oxygen-rich growth conditions, the oxygen chemical potential 
is $\mu_\mathrm{O}=0$.
In the titanium-rich (oxygen-poor) limit,
$\mu_\mathrm{O}$ is determined by the formation of Ti$_2$O$_3$,
which implies the condition 
$2\mu_{\mathrm{Ti}}+3\mu_{\mathrm{O}}=
\Delta{H}_{\mathrm{f}}(\mathrm{Ti}_2\mathrm{O}_3)$.
Here $\Delta{H}_{\mathrm{f}}(\mathrm{Ti}_2\mathrm{O}_3)$ is the formation
enthalpy of Ti$_2$O$_3$, which we found to be $-15.33$~eV (per formula unit).
On the other hand, stability condition for the \tio\ requires
$\mu_{\mathrm{Ti}}+2\mu_{\mathrm{O}}=\Delta{H}_{\mathrm{f}}(\mathrm{TiO}_2)$,
where formation enthalpy $\Delta{H}_{\mathrm{f}}(\mathrm{TiO}_2)=-9.66$~eV
(per formula unit). From these two conditions we find the oxygen chemical potential
to be $\mu_{\mathrm{O}}=-3.99$~eV in the titanium-rich limit.

As can be seen from Fig.~\ref{fig:form_en}, the most stable defects in the wide range
of possible values for Fermi energy are $+2$ charged oxygen vacancies.
This finding is in qualitative agreement with the previous HSE study by
Janotti {\em et al.} [see Fig.~5 of Ref.~\onlinecite{janotti_10}].
Similar to that work, we also find
that the transition from the $+2$ to neutral state occurs at a higher value
of $E_{\mathrm{F}}$ than the transition from the neutral to $+1$ state
(a feature of the negative $U$ defect). Quantitatively, however,
our values for charge transition levels are smaller then what was found in 
Ref.~\onlinecite{janotti_10} by $\sim0.5$~eV. To be more precise, charge
transition levels in that work were found to be at or above the conduction
band minimum and, as a result, the $V_{\mathrm{O}}^{+2}$ was found to be the 
only stable oxygen vacancy for all values of $E_{\mathrm{F}}$. In our case,
we find that for $E_{\mathrm{F}}>2.8$~eV the neutral vacancy can become 
more stable.

We emphasize that the study of formation energies 
and relative stability of charged oxygen vacancies in rutile \tio\ cannot be done 
at the PBE level since in this case the defect levels are not found in the bulk band gap.
Therefore, it is crucial to use more advanced methods, such as, e.g., the one described above.

\begin{table}
\caption{\label{tab:ctl}
Contributions to the charge transition levels coming from the
quasiparticle energy $E_{\mathrm{QP}}$, relaxation energy
$E_{\mathrm{relax}}$, and electrostatic correction
$\Delta{E}^{\mathrm{e.s.}}_{\mathrm{QP}}$ 
(all values are given in eV).}
\begin{ruledtabular}
\begin{tabular}{lcc}
  & $\epsilon^{2+/1+}$   & $\epsilon^{1+/0}$ \\
\hline
$E_{\mathrm{QP}}$    & \phm{$-$}2.58 & \phm{$-$}3.24 \\
$E_{\mathrm{relax}}$ & \phm{$-$}0.86 &       $-$0.29 \\
$\Delta{E}^{\mathrm{e.s.}}_{\mathrm{QP}}$ & $-$0.44 &       $-$0.44 \\
\hline
$\epsilon^{q/q-1}$   & \phm{$-$}3.00 & \phm{$-$}2.51 \\ 
\end{tabular}
\end{ruledtabular}
\end{table}

\section{Summary}
\label{sec:summary}

In summary, we investigated the oxygen vacancies in rutile \tio\ in three
charge states from first principles using the DFT+GW approach. 
The oxygen vacancies were emulated
in a 71 atom supercells. The structural relaxations around the defects
were performed using the hybrid functional (HSE) method and
charge transition levels and defect formation energies were calculated 
within the DFT+GW formalism. According to our calculations, in a wide range
of values for Fermi energy, $0<E_{\mathrm{F}}<2.8$~eV, the $+2$ charge state
of the vacancy is the most stable, while for Fermi energies above $2.8$~eV
the neutral vacancy is stabilized. This result also means that oxygen vacancy
is found to be a negative $U$ defect. 

\section{Acknowledgments}
\label{sec:acknowledgments}

This work was supported by National Science Foundation Grant No. DMR10-1006184
(ground-state and structural studies, electrostatic correction analyses,
and effective mass calculations)
and the Theory Program at the
Lawrence Berkeley National Laboratory (LBNL) funded by the Department of Energy (DOE),
Office of Basic Energy Sciences, under Contract No. DE-AC02-05CH11231 (quasiparticle 
calculations and studies of charge transition levels).  Algorithm developments for large-scale
GW simulations were supported through the Scientific Discovery through Advanced Computing 
(SciDAC) Program on Excited State Phenomena in Energy Materials funded by DOE, Office 
of Basic Energy Sciences and of Advanced Scientific Computing Research, under Contract 
No. DE-AC02-05CH11231 at LBNL. SGL acknowledges support of a Simons Foundation 
Fellowship in Theoretical Physics.  Computational resources have been provided by DOE 
at Lawrence Berkeley National Laboratory’s NERSC facility and by National Institute for 
Computational Sciences.

We would like to thank A. Janotti for helpful discussions.

\pagebreak
\widetext
\begin{center}
\textbf{\large Supplemental Material to ``First-principles DFT+GW study of oxygen vacancies in rutile \tio''}
\end{center}

\setcounter{equation}{0}
\setcounter{figure}{0}
\setcounter{table}{0}
\setcounter{page}{1}
\makeatletter
\renewcommand{\theequation}{S\arabic{equation}}
\renewcommand{\thefigure}{S\arabic{figure}}
\renewcommand{\bibnumfmt}[1]{[S#1]}
\renewcommand{\citenumfont}[1]{S#1}
\section{Introduction}
\label{app_sec:intro}

  In this Supllemental Material we provide the details of calculation of the band structure and 
quasiparticle effective masses of rutile
  \tio\ from first principles within the GW formalism. For quasiparticle 
  eigenvalues, we use a G$_0$W$_0$ method based on the 
  Hybertsen-Louie generalized plasmon pole model. Within this model,
  the convergence of the self-energy is assured by including
  a sufficient number of conduction bands in the calculation. To further verify
  the accuracy of the calculations, we perform an additional full-frequency
  G$_0$W$_0$ calculation of the direct band gap at the $\Gamma$ point.
  We found the fundamental band gap in rutile to be an indirect $\Gamma-\mathrm{R}$
  gap of $3.13$~eV.
  The quasiparticle effective masses are 
  computed for the quasielectrons of the lowest conduction band and 
  quasiholes of the highest valence band using Wannier interpolation technique.

The precise knowledge of the electronic band structure of rutile \tio\ 
is crucial for understanding its optical properties. Despite enormous research
efforts, both theoretical and experimental, controversies still remain
in the evaluation of such basic quantities as quasiparticle band gaps for this material.

Early on rutile was found to have an optical band gap of $3.05$~eV from
optical absorption and photoconductivity measurements.
\cite{app_cronemeyer_51,app_cronemeyer_52,app_cronemeyer_59}
Later measurements\cite{app_arntz_66,app_vos_74,app_pascual_77,app_pascual_78,app_amtout_95}
resolved the fine structure of the absorption edge in rutile
showing that the first peak in the absorption spectrum appears at $3.031$~eV.
On the other hand, photoemission and inverse photoemission experiments
have shown that the electronic band gap, defined as the difference between
the conduction-band minimum (CBM) and valence-band maximum (VBM), is higher
than the optical gap. The reported values for the electronic band gap vary
in the range $3.3-4.0$~eV\cite{app_hardman_94,app_tezuka_94,app_rangan_10}
with one of the most accurate values reported recently being
$3.6\pm0.2$~eV.\cite{app_rangan_10} The optical properties of anatase and brookite
are less studied compared to rutile. The reported values
for the optical absorption gaps for anatase and brookite are
$\sim3.4-3.8$~eV\cite{app_tang_95,app_wang_02} and $\sim3.3$~eV, respectively.
The electronic band gaps have to be larger than the corresponding optical gaps
because of exciton formation.
However, to the best of our knowledge,
they have not yet been directly measured in these phases.

As for theory, numerous first-principles calculations have been done
in the past years. An extensive comparison of values of band gaps for rutile
and anatase obtained with different theoretical methods can be found 
in Ref.~\onlinecite{app_chiodo_10}. Typically, mean-field calculations
based on density-functional theory (DFT) underestimate the band gap 
substantially. This is a well known problem, which takes its roots
from the fact that the Kohn-Sham eigenvalues do not represent actual
quasiparticle energies. One of the most successful approaches
to mitigate this problem is based on the many-body perturbation
theory, employing the so-called GW method.~\cite{app_hybertsen_86}
However, the reported values for band gaps in \tio\ obtained 
with the help of GW method still vary significantly. For example,
in the case of rutile the values were reported from $2.9$~eV
to $4.8$~eV.\cite{app_chiodo_10,app_patrick_12,app_kang_10} 
There could be several reasons for this broad range of values.
On one hand, the GW method itself has many flavors with different level 
of approximation. It is a subject
of many debates regarding to which flavor is appropriate. On the other hand,
a typical GW calculation is much more computationally demanding
than the corresponding mean-field calculation. It requires
a more careful convergence with respect to a larger set
of parameters. A fully converged
GW calculation is a challenging task but necessary for accurate results.
In particular, it has been shown recently that wurtzite ZnO
requires several thousands of empty bands 
to converge the GW band gap calculation.\cite{app_shih_10,app_friedrich_11}
Here, we revisit the problem of the band gap in rutile \tio.
We perform a GW calculation of the band structure,
making sure that convergence with respect to number of bands has been
achieved.
Based on the band structure data,
we also perform calculations of the first effective masses of quasiparticles 
for specific extremal points of the highest valence 
and lowest conduction bands.

The rest of the Supplemental Material is organized as follows.
Section~\ref{app_sec:computational} describes computational details
of our DFT and GW calculations. Results are discussed
in Sec.~\ref{app_sec:results} followed by a summary
in Sec.~\ref{app_sec:summary}.

\section{Computational details}
\label{app_sec:computational}

In this work, mean-field calculations are carried out using a plane-wave
{\it ab initio} pseudopotential approach to DFT as implemented
within the {\sc Quantum ESPRESSO} code package.\cite{app_QE_09}
We used generalized-gradient 
approximation (GGA) with the Perdew-Burke-Ernzerhof (PBE)
parameterization\cite{app_perdew_96}
of the exchange-correlation energy functional.
Troullier-Martins norm-conserving pseudopotentials\cite{app_troullier-prb91}
were generated with Ti $3s$ and $3p$
states treated as valence orbitals.
The titanium pseudopotential was generated in the
Ti$^{4+}$ configuration, and the cut-off radii for the $3s$, $3p$,
and $3d$ states were chosen
to be $0.9$, $0.9$, and $1.0$ a.u., respectively.

The plane-wave basis was determined by an energy cutoff of 200~Ry.
The Brillouin zone was sampled by a uniform $4\times4\times6$ $k$-point mesh.
These parameters were sufficient to obtain a well converged mean-field
calculation for the band structure. 
E.g., the value of the PBE direct band gap at the $\Gamma$ point
increases by about $0.01$~eV when the $k$-point mesh is changed
to a denser $6\times6\times9$ mesh.
(The corresponding GW value changes by $0.04$~eV.)

The structural parameters of bulk rutile were determined theoretically.
Since the band structure results presented
in this Supplemental Material are obtained using the GW method with 
the starting PBE mean field, for consistency we used PBE structural parameters
(see Table I of the main text).

\begin{figure}
\centering\includegraphics[width=246pt]{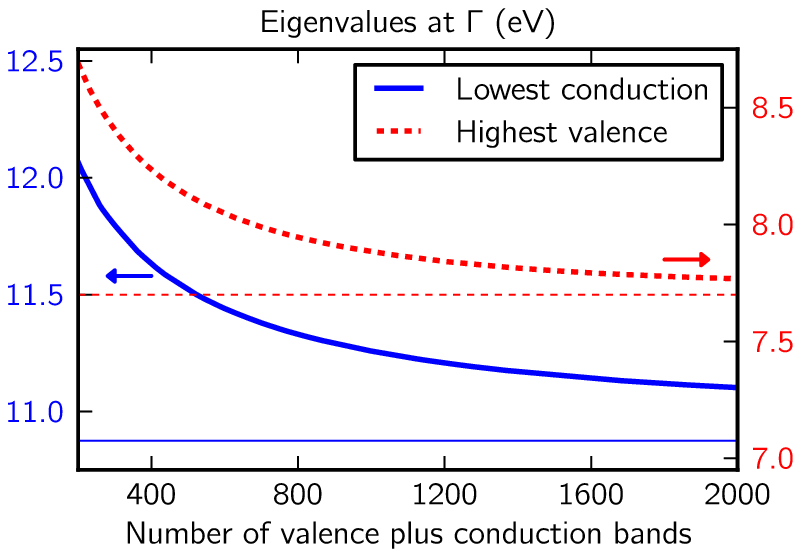}
\caption{(Color online.) 
  Dependence of the highest valence and lowest conduction 
  quasiparticle eigenvalues at the $\Gamma$ point 
  on the total number of bands used in the self-energy calculation.
  The horizontal thin lines show the converged values obtained with the static 
  remainder method.
}
\label{fig:conv}
\end{figure}

The GW calculations were performed using the
BerkeleyGW code package.\cite{app_deslippe_12}
To ensure convergence with respect to number of empty states in the 
self-energy calculation, we used $2\,000$ valence plus conduction bands.
This number of bands ensures that we include all states within
approximately 40~Ry above the Fermi level.
In addition, the static remainder method\cite{app_deslippe_13}
was used in order to 
confirm that convergence has indeed been achieved.
As an example, the dependence of the computed highest valence 
and lowest conduction eigenvalues at the $\Gamma$ point 
on the number of bands
is shown in Fig.~\ref{fig:conv}. One can see from the figure
that convergence of the quasiparticle eigenvalues with respect
to empty states is rather slow. On the other hand, if one is interested
only in the calculation of the differences between the eigenvalues
(band gaps, e.g.), the convergence may be achieved faster.
E.g., as cen be seen from Fig.~\ref{fig:conv}, the value of the direct band gap
computed with 200 bands is only about $0.2$~eV larger then the converged value.

The majority of our 
GW calculations were performed at the G$_0$W$_0$ level employing
the generalized plasmon pole (GPP) model and
full frequency G$_0$W$_0$ method
was used at the $\Gamma$ point as an additional check.

For the calculation of the GW quasiparticle band structure 
the following procedure was used.
First, the PBE Kohn-Sham eigenstates and eigenvalues were obtained
on a $4\times4\times6$ grid of $k$ points. Then the eigenvalues were corrected
by employing a G$_0$W$_0$ approach within the complex 
GPP model.\cite{app_zhang_89}
Finally, Wannier interpolation scheme\cite{app_mostofi_08,app_marzari_12} 
was employed in order to obtain a fine set of eigenvalues along 
the high-symmetry directions
in the Brillouin zone. Within this scheme, the maximally localized 
Wannier functions\cite{app_marzari_97} were constructed
from the mean-field (PBE) wavefunctions while GW eigenvalues were used
at the band structure interpolation step.
  The accuracy of the Brillouin zone interpolation was checked
  by doing interpolation starting from $4\times4\times6$ and $6\times6\times9$
  coarse $k$-point grids and comparing the values of effective masses
  (see next section for details on effective mass calculations).
  To reduce the computational cost, we did this comparison only at the PBE level.
  We found that to first two significant digits the values of effective
  masses did not change, except at the $\Gamma$ point. Therefore, 
  effective masses at this point were calculated separately, without
  invoking the interpolation procedure.

\section{Results}
\label{app_sec:results}

As for the band gap, at the PBE level we found the fundamental gap to be
a direct gap of $1.86$~eV at the $\Gamma$ point.
At the G$_0$W$_0$ level, the gap at $\Gamma$ increased to $3.18$~eV and 
the fundamental gap became an indirect $\Gamma-\mathrm{R}$ gap of $3.13$~eV,
which is still very close to the value of the direct gap at $\Gamma$.
While the valence-band maximum clearly occurs at the $\Gamma$ point, 
the minima of the lowest conduction band at $\Gamma$, $\mathrm{M}$,
and $\mathrm{R}$ points
are very close. This finding is consistent with previous GW
calculations.\cite{app_kang_10,app_chiodo_10} 
Kang and Hybertsen\cite{app_kang_10} also found the fundamental
gap to be an indirect $\Gamma-\mathrm{R}$ one with a slightly higher value
of $3.34$~eV. While Chiodo {\it et al.}\cite{app_chiodo_10} reported the direct gap
at $\Gamma$ to be the lowest one, Fig.~$2$ in their paper shows that the lowest GW gap
is in fact also $\Gamma-\mathrm{R}$. In any case, the band structure
of rutile \tio\ is found to be rather peculiar, with a direct band gap at $\Gamma$
being very close to the indirect $\Gamma-\mathrm{R}$ and $\Gamma-\mathrm{M}$ gaps.

As discussed in the main text, to study the charged oxygen vacancies in rutile,
we had to use hybrid HSE functional for structural relaxations. It is natural
to ask then how the value of the GW band gap in rutile depends on the structural 
parameters of the system. Starting from the HSE parameters shown in Table~I
of the main text, we performed PBE mean field calculations and applied the GW correction.
We found the direct gap at $\Gamma$ in this case to be $3.25$~eV, very close to 
the value of $3.18$~eV obtained with the PBE structural parameters.
Thus, for the purpose of calculation of defect formation energies and charge transition
levels, this difference is clearly insignificant, given the
number of approximations made (e.g., in the evaluation of electrostatic corrections).

It is also important to check the robustness of GW results with respect to
the starting mean field calculation. We performed a one-shot G$_0$W$_0$
calculation of the direct gap at $\Gamma$ starting from mean field
obtained with local-density approximation (LDA) exchange-correlation
functional (using the same HSE structural parameters as above).
We found the gap in this case to be $3.28$~eV, very close to the GW value of
$3.25$~eV obtained with the PBE reference mean field. For comparison, the LDA
gap is $1.77$~eV and the PBE is $1.90$~eV. Thus, the GW correction
depends on the reference DFT calculation and adjusts itself in such a way
as to give very close final GW values.

In order to assess the accuracy of our GPP G$_0$W$_0$ results, we
also carried out a full-frequency G$_0$W$_0$ calculation. Since this type 
of calculations is much more computationally demanding than the
GPP G$_0$W$_0$ calculations, we decided to do a full-frequency
integration only at the $\Gamma$ point. The frequency integration
was done along the real axis\cite{app_deslippe_12} using
regular mesh of frequencies up to $50$~eV with spacing of $0.2$~eV and
then using linearly increasing spacing for frequencies up to cutoff of
$1\,000$~eV. The direct band gap at $\Gamma$ computed in this fashion
was found to be $3.16$~eV, in excellent agreement with our GPP
G$_0$W$_0$ result. 

Using the band-structure data shown in Fig.~2 of the main text,
one can determine the effective masses of the quasiparticles
at the band extrema. Here we analyze highest valence 
and lowest conduction bands. 
Of special interest are the quasiparticles associated with the 
$\Gamma$, $\mathrm{R}$, and $\mathrm{M}$ points of the Brillouin zone since
the lowest conduction band has almost the same energy at these points
with conduction band minimum being at $\mathrm{R}$. The point $\mathrm{A}$ is also
of interest due to a local maximum of the valence band. 
Table~\ref{tab:m_eff} lists the effective masses calculated at these four points.

Note that tetragonal symmetry of rutile implies that the main axes
of the effective mass tensor at $\Gamma$ coincide with the
Cartesian axes $k_x$, $k_y$, and $k_z$ [see Fig.~1(b) of the main text],
with $k_x$ and $k_y$ being equivalent. Therefore,
the effective mass tensor at this point must be diagonal in $k_x-k_y$ plane
and the effective masses in the columns $\Gamma\ra\mathrm{X}$ and $\Gamma\ra\mathrm{M}$
in Table~\ref{tab:m_eff} must be the same. 
One can see from the Table that apart from small numerical
noise this is indeed the case.

The effective mass tensors (both for the highest valence and lowest conduction
bands) of rutile at $\Gamma$ is highly anisotropic.
Quasi-electrons are less massive by a factor of $\sim3$ in response to a perturbation
along $z$ direction compared to a response perpendicular to $z$ axis.
Quasi-holes at $\Gamma$, however, are more massive along $z$ direction.
They are also heavier compared to quasi-electrons.
Interestingly, compared to other band minima, quasi-electrons at $\Gamma$ 
are both the heaviest and the lightest depending on the direction of response.

The computed effective masses are in reasonable agreement with previous
$\sigma$-GGA and $\sigma$-GGA$+U$ calculations,\cite{app_perevalov_11}
except for the $\Gamma\ra(\mathrm{X},\mathrm{M})$
electron effective mass, which in our case is an order of magnitude smaller than 
the $\sigma$-GGA$+U$ value.
The experimental values for the electron effective masses
are reported in a wide range, from $\sim0.7$~$m_e$\cite{app_stamate_03} to 
$\sim3$~$m_e$,\cite{app_pascual_77,app_pascual_78} where $m_e$
is the electron mass in vacuum.
The wide range of reported values may be partially attributed to the polaron
effects\cite{app_pascual_78} in \tio\ and associated difficulties in extracting 
the bare effective mass from the polaron effective mass in this case. 
The polaron effects on the effective masses are not considered in the present work.

\begin{table*}
  \caption{\label{tab:m_eff}
    Calculated effective masses for rutile \tio\ 
    expressed in units of the electron mass, $m_e$. Negative sign indicates holes.
Notation $P_1\ra{P_2}$ specifies that effective mass was computed 
from the band curvature in the vicinity of point $P_1$ in direction
towards point $P_2$.}
  \begin{ruledtabular}
    \begin{tabular}{lccccccc}
     PBE & $\Gamma\ra\mathrm{X}$ & $\Gamma\ra\mathrm{M}$ &
          $\Gamma\ra\mathrm{Z}$ & $\mathrm{R}\ra\mathrm{X}$ &
          $\mathrm{R}\ra\mathrm{Z}$ & $\mathrm{A}\ra\mathrm{M}$ &
          $\mathrm{M}\ra\mathrm{A}$ \\ 
      Conduction band & $\phm{-}1.2$ & $\phm{-}1.2$ & $\phm{-}0.6$ & $\phm{-}1.0$ &
                   $\phm{-}0.6$ & $\phm{-}1.1$ & $\phm{-}0.8$ \\ 
      Valence band    &       $-3.1$ &       $-3.1$ &       $-5.3$ &       $-3.0$ &
                    $-0.5$ &       $-1.6$ &       $-1.2$ \\
      \hline
     GW & $\Gamma\ra\mathrm{X}$ & $\Gamma\ra\mathrm{M}$ &
          $\Gamma\ra\mathrm{Z}$ & $\mathrm{R}\ra\mathrm{X}$ &
          $\mathrm{R}\ra\mathrm{Z}$ & $\mathrm{A}\ra\mathrm{M}$ &
          $\mathrm{M}\ra\mathrm{A}$ \\ 
      Conduction band & $\phm{-}1.4$ & $\phm{-}1.5$ & $\phm{-}0.6$ & $\phm{-}1.0$ &
                   $\phm{-}0.7$ & $\phm{-}1.1$ & $\phm{-}0.8$ \\ 
      Valence band    &       $-2.4$ &       $-2.4$ &       $-3.6$ &       $-3.5$ &
                    $-0.6$ &       $-1.4$ &       $-1.0$
    \end{tabular}
  \end{ruledtabular}
\end{table*}

\section{Summary}
\label{app_sec:summary}

We performed a theoretical study of the basic electronic structure properties,
such as the quasiparticle band structure and effective masses,
of rutile \tio\ by means of first-principles calculations based on
the GPP G$_0$W$_0$ method. Strict convergence criteria required us
to use about $2\,000$ conduction bands in the evaluation of the self-energy.
The accuracy of the method was cross checked by performing 
an additional full frequency G$_0$W$_0$ calculation. Our band gap results for
rutile are in good agreement with previous GW and experimental studies.
In particular, we found the fundamental gap to be an indirect $\Gamma-\mathrm{R}$
one. The value of the gap was found to be $3.13$~eV.
The quasiparticle eigenvalues were evaluated on a mesh of $k$ points
and then interpolated to high symmetry lines in the Brillouin zone
allowing us to evaluate the effective masses.
The quasiparticle effective masses were computed for certain local extrema
of the lowest conduction and highest valence bands. We found the effective masses
to be highly anisotropic for both quasielectrons and quasiholes. The quasielectron
effective masses vary from $0.6$~$m_e$ to $1.5$~$m_e$, while the quasiholes
are generally heavier and have masses from $0.6$~$m_e$ to $3.6$~$m_e$
depending on direction.

\end{document}